\documentclass[conference]{IEEEtran}
\IEEEoverridecommandlockouts
\usepackage{cite}
\usepackage{amsmath,amssymb,amsfonts}
\usepackage{graphicx}
\usepackage{textcomp}
\usepackage{xcolor}

\usepackage[hyphens]{url}
\usepackage{fancyhdr}
\usepackage[bookmarks=true,breaklinks=true,letterpaper=true,colorlinks,citecolor=blue,linkcolor=blue,urlcolor=blue]{hyperref}

\usepackage{subfiles}
\usepackage{diagbox}
\usepackage{booktabs}
\usepackage{xspace}
\usepackage{tabularx}
\usepackage{wrapfig}  
\usepackage{bbding}
\usepackage{pifont}
\usepackage{threeparttable}

\usepackage{multirow}
\usepackage[capitalize,noabbrev]{cleveref}
\usepackage{mathrsfs}
\usepackage[ruled,vlined,linesnumbered ]{algorithm2e} 
\usepackage{placeins}

\def\BibTeX{{\rm B\kern-.05em{\sc i\kern-.025em b}\kern-.08em
    T\kern-.1667em\lower.7ex\hbox{E}\kern-.125emX}}
\begin{document}

\newcommand{\NAME}{APACHE\xspace} 
\newcommand{\data}{DHO\xspace} 
\newcommand{\computation}{CHO\xspace} 
\newcommand{\pmult}{\mathsf{PMult}}
\newcommand{\cmult}{\mathsf{CMult}}
\newcommand{\lwe}{\mathsf{LWE}}
\newcommand{\rlwe}{\mathsf{RLWE}}
\newcommand{\rgsw}{\mathsf{RGSW}}
\newcommand{\privks}{\mathsf{PrivKS}}
\newcommand{\pubks}{\mathsf{PubKS}}
\newcommand{\decomp}{\mathsf{Decomp}}
\newcommand{\nintt}{\mathsf{(I)NTT}}
\newcommand{\mmult}{\mathsf{MMult}}
\newcommand{\madd}{\mathsf{MAdd}}
\newcommand{\hadd}{\mathsf{HAdd}}
\newcommand{\auto}{\mathsf{Automorph}}
\newcommand{\hrot}{\mathsf{HRot}}
\newcommand{\ntt}{\mathsf{NTT}}
\newcommand{\intt}{\mathsf{INTT}}
\newcommand{\bconv}{\mathsf{BConv}}
\newcommand{\keyswith}{\mathsf{KeySwith}}

\title{\NAME: A Processing-Near-Memory Architecture for 
Multi-Scheme Fully Homomorphic Encryption}
\author{Lin Ding, Song Bian, Penggao He, Yan Xu, Gang Qu, and Jiliang Zhang~\thanks{Corresponding author: Jiliang Zhang}}

\maketitle

%


\begin{abstract}
Fully Homomorphic Encryption (FHE) is known to be extremely computationally-intensive, application-specific accelerators emerged as a powerful solution to narrow the performance gap. Nonetheless, due to the increasing complexities in FHE schemes per se and multi-scheme FHE algorithm designs in end-to-end privacy-preserving tasks, existing FHE accelerators often face the challenges of low hardware utilization rates and insufficient memory bandwidth. In this work, we present \NAME, a layered near-memory computing hierarchy tailored for multi-scheme FHE acceleration. By closely inspecting the data flow across different FHE schemes, we propose a layered near-memory computing architecture with fine-grained functional unit design to significantly enhance the utilization rates of computational resources and memory bandwidth. 
The experimental results illustrate that \NAME outperforms state-of-the-art ASIC FHE accelerators by 10.63$\times$ to 35.47$\times$ over a variety of application benchmarks, e.g., Lola MNIST, HELR, VSP, and HE$^{3}$DB.
\end{abstract}


\section{Introduction}
\label{section: Introduction}

With the increasing popularity of cloud-based storage and computation outsourcing, fully homomorphic encryption (FHE)~\cite{FHESurvey} rises as one of the most promising solutions to secure data confidentiality while maintaining its usability. Modern FHE allows complicated algorithms to be directly applied to ciphertexts, without revealing the decryption key to service providers. 
Recently, increasing attention focuses on designing protocols using multi-scheme FHE constructions for enhanced flexibility. In such a case, multiple FHE schemes, e.g., TFHE \cite{tfhe} and CKKS~\cite{CKKS}, are used to (jointly or separately) establish privacy-preserving protocols, such as private machine learning \cite{LR_AAAI2019,Lola}, fully homomorphic processors \cite{VSP}, private medicine \cite{CAV2022}, and encrypted database \cite{HE3DA}. Among them, schemes such as CKKS and BFV \cite{BFV} are often used to evaluate polynomial functions (e.g., additions, multiplications), whereas TFHE-like schemes \cite{FHEW,tfhe} are employed to evaluate non-polynomial functions (e.g., logic operations).

\begin{figure}[t]
    \centering
    \includegraphics[width=0.95\linewidth]{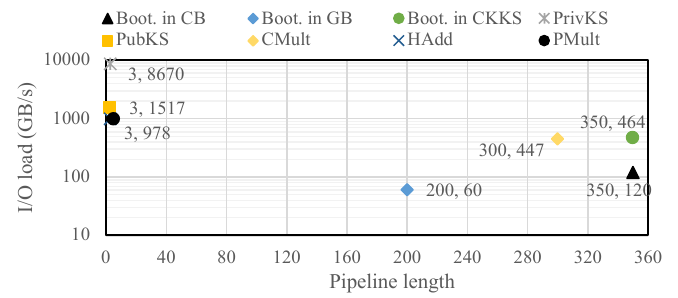}
    \vspace{-12 pt}
    \caption{Evaluation of I/O load in the pipelined accelerator, referring to \cite{tfhe,HE3DA,Poseidon}.}
    \label{fig:IO load}
    \vspace{-12 pt}
\end{figure}

While protocols based on multi-scheme FHE can be much more expressive and usable,  we face new computational and communicational obstacles against the associated architectural designs. As illustrated in \cref{fig:IO load}, we first categorize basic FHE algorithms into data-heavy operators and computation-heavy operators. Multi-scheme-FHE-based protocols need to include a much wider range of fundamental FHE operators to support complex computations across different privacy-preserving tasks. However, as also sketched in \cref{fig:IO load}, due to varying ciphertexts parameters and evaluation key sizes (e.g., bootstrapping keys, key-switching keys, rotation keys), FHE operators of different schemes exhibit distinct data-heavy and computation-heavy characteristics, necessitating new explorations in the architectural design space of FHE accelerators. 

Based on the above analyses, we identify three key design principles for multi-scheme FHE accelerators: 1) layered memory access to reduce I/O load, 2) reconfigurable functional units for flexible operator granularity, and 3) operator scheduling to maximize data parallelism. Using such critiques, we evaluated existing FHE accelerators in \cref{table:compaBetacc} and reveal two insights. First, prior accelerators often rely on a two-level memory hierarchy with off-chip memory at low bandwidth and of-chip memory at high bandwidth. Accelerators, like \cite{Alchemist,F1, CraterLake,ARK, BTS}, use High-Bandwidth Memory (HBM) for key and ciphertext buffering, providing up to 2 TB/s I/O. However, \cref{fig:IO load} notes that 8 TB/s is needed for full utilization in circuit bootstrapping units of TFHE. 
Second, current architectures have fixed interconnects for specific schemes, lacking effective task scheduling, which lowers resource utilization. For instance, \cite{F1,CraterLake,BTS,SHARP_ISCA2023} includes a BConv unit for BFV/CKKS evaluation, leaving 30\% of chip area, but the BConv operation is unused in TFHE. Hence, a natural question is that, how can we co-design the memory and computation architectures to support multi-scheme FHE operators, where both the memory bandwidth and computing resources can be fully exploited?

\begin{scriptsize}
\footnotesize
\begin{table*}[t]
  \centering
  \caption{Qualitative comparisons between homomorphic accelerators.}
  \label{table:compaBetacc}
  \vspace{-8 pt}
  \begin{tabular}{lccccccccccccccccc}
  \toprule  
  &\cite{F1}&\cite{Strix}&\cite{CraterLake}&\cite{ARK}&\cite{BTS}&\cite{Poseidon}&\cite{INSPIRE}&\cite{MATCHA}&\cite{FPT}&\cite{Poseidon-NDP}&\cite{TVLSI2014}&\cite{CHES2018}&\cite{CHES2021}&\cite{Alchemist}&\cite{HEAP}&Ours\\
   \midrule
   
   TFHE-like$^a$ &$\vartriangle$&\ding{52}&\ding{56}&\ding{56}&\ding{56}&$\vartriangle$&\ding{56}&\ding{52}&\ding{52}&$\vartriangle$&\ding{56}&\ding{56}&\ding{56}&\ding{52}&$\vartriangle$&\ding{52}\\
   
   I/O bus load &High&Low&High&High&High&High&Low&High&Low&Low&High&High&High&High&High&Low\\
    Configurability&\ding{56}&\ding{52}&\ding{56}&\ding{56}&\ding{56}&\ding{56}&\ding{56}&\ding{56}&\ding{56}&\ding{56}&\ding{56}&\ding{52}&\ding{52}&\ding{52}&\ding{52}&\ding{52}\\
    
   Parallelism &$\blacktriangle$&$\blacktriangle$&$\blacktriangle$&$\blacktriangle$&$\blacktriangle$&\ding{56}&$\blacktriangle$&$\blacktriangle$&\ding{56}&\ding{56}&\ding{56}&$\blacktriangle$&$\blacktriangle$&$\blacktriangle$&\ding{52}&\ding{52}\\
  \bottomrule
  \end{tabular}
  \begin{tablenotes}
  \item [] $\blacktriangle$ denotes that a accelerator has paralleled multiple cores but does not support the parallelism at the level of accelerators; $^a$\cite{MATCHA,FPT,Strix,Alchemist} only support the gate bootstrapping function of TFHE, while the proposed \NAME is capable of implementing both of gate bootstrapping and circuit bootstrapping functions. $\vartriangle$ marks such works that claim to support at least one logic FHE scheme but without any evidence and description of the implementation. 
  \end{tablenotes}
  \vspace{-11 pt}
\end{table*}
\end{scriptsize}

{\bf{Our Contribution}}: In this work, we propose \NAME, a processing-near-memory (PNM) architecture designed for general-purpose FHE acceleration. First, we leverage a three-level memory hierarchy to fully exploit the internal bandwidth of the DIMM 
units~\cite{Near-Memory_TCASI2022,Triangle_TC2022}. By lifting the large-volume communication from external I/O loads, we are able to avoid mounting an excessive amount of expensive HBM units. Moreover, we also leverage task-level operator parallelism to maximize the parallel capabilities of multi-channel DIMMs. Meanwhile, we propose a new circuit topology to interconnect functional units (FU), such that a high hardware utilization rate can be retained across the diverse set of FHE operators. The key contributions of our work can be summarized as follows.

\begin{itemize}
    \item {\bf{Multi-Level PNM Architecture}}: 
    Our key observation is that data-heavy operators generally have a low computational circuit depth, as \cref{fig:IO load} shows. Thus, we allocate functional units into three memory levels: the external I/O level, the near-memory level, and the in-memory level. By placing computation-light but memory-heavy units closer (or even into) the memory die, we reduce the I/O bandwidth by up to 3.15 $\times$ $10^5$ times. 
    \item {\bf{Configurable Internconnect Topology}}: We propose a flexible interconnect topology with detailed FU designs to support multi-scheme operators and optimize hardware utilization. \NAME shows that (I)NTT FU utilization remains above 90\% for multi-scheme FHE tasks, while that of existing schemes ranges from 50\% to 85\%.
    \item {\bf{Implementation and Evaluation}}: 
    We thoroughly evaluate the performance of \NAME on a set of multi-scheme FHE operator and application benchmarks. Results show that \NAME achieves 4.07$\times$ to 35.47$\times$ speedup over state-of-the-art FHE accelerators.
\end{itemize}

The rest part of this paper is organized as follows. In \cref{section: Background and Motivation}, we present the background and related works. Then, we overview \NAME architecture in \cref{section: Architecture}. Next, we show the configurable underlying circuits and memory layouts in \cref{section: Functional Unit}. Subsequently, we provide the experimental results in \cref{section: Implementation}. Finally, we conclude our work in \cref{section: Conclusion}.


\begin{table*}[t]
    \footnotesize
    \centering
    \caption{Decomposition and classification of homomorphic operators.}
    \vspace{-8 pt}
    \begin{tabular}{l|l|cccc|lccc|c}
    \toprule
    \multicolumn{2}{c|}{\multirow{2}{*}{Homomorphic Operators}} & \multicolumn{4}{c|}{Basic Functional Units} & \multicolumn{1}{c}{Pipeline} & \multicolumn{1}{c}{Cached}& \multicolumn{1}{c}{Operand} & \multicolumn{1}{l|}{Input} & Operator \\
    \cline{3-6}
    \multicolumn{2}{c|}{ } & NTT & MA & MM & Auto. & 
    \multicolumn{1}{c}{Depth$^1$} & \multicolumn{1}{c}{Key Size}&  \multicolumn{1}{c}{Bitwidth$^2$} & \multicolumn{1}{c|}{Sym.$^3$} & Type \\ \midrule
    \multirow{5}{*}{\shortstack{TFHE-like \\\cite{VSP,ASIACCS2023}}} & $\mathsf{CMUX}$ & \ding{52} &\ding{52} &\ding{52} &\ding{52}& $\leq$ 350 & None & 32, 64 & \ding{52} & Computation\\
     & $\mathsf{PrivKS}$ & &\ding{52} &\ding{52} & & $\leq$ 3 & 1.8 GB & 64 & \ding{56} & Data \\
     & $\mathsf{PubKS}$ & &\ding{52}&\ding{52}& &$\leq$ 3 & 79 MB & 32 & \ding{56} & Data\\
     & Gate Boot.$^4$ &\ding{52}&\ding{52}&\ding{52}&\ding{52}& $\leq$ 350 & 37 MB & 32 & \ding{52} & Computation\\
     & Circuit Boot.$^4$ &\ding{52}&\ding{52}&\ding{52}&\ding{52} & $\leq$ 350 & 196 MB & 32, 64 &\ding{52} & Computation\\
    \midrule
    \multirow{3}{*}{\shortstack{BFV- and \\CKKS-like \\\cite{ARK}}} 
    & $\mathsf{HAdd}$ & &\ding{52}& & & $\leq$ 3 & None & $\leq$ 32 & \ding{52} & Data\\
     & $\mathsf{HMult}$ & \ding{52} &\ding{52} &\ding{52} & & $\leq$ 300 & 120 MB & $\leq$ 32 & \ding{52} & Computation\\
    & CKKS Boot. & \ding{52} &\ding{52} &\ding{52} &\ding{52} & $\leq$ 350 & $\approx$ 1 GB & $\leq$ 32 & \ding{52} & Both\\
    \bottomrule     
    \end{tabular}
    \begin{tablenotes}
	\item[] $^1$Commonly, a full-pipelined NTT unit is comprised of 
 150 to 250 stages of circuits. The pipeline lengths of MA and MM are 
 generally less than three and five, respectively. As designed in~\cite{F1}, 
 an 128-lane automorphism module needs at least 63 stages of circuits. The 
 overall circuit depth is estimated based on the above parameters.
 $^2$Operands of 128 bits or above are typically split into integers of 32 bits or less for efficient calculation. 
 $^3$Input Symmetry means that all the input operands have the 
 same bitwidth. 
 $^4$Excluding the built-in key switching procedures. $\mathsf{HomGate}$ is also known as Gate Boot.
	\end{tablenotes}
    \label{tab:Classification_OP}
    \vspace{-8 pt}
\end{table*}

\begin{figure*}[t]
    \footnotesize
    \centering
    \includegraphics[width=\linewidth]{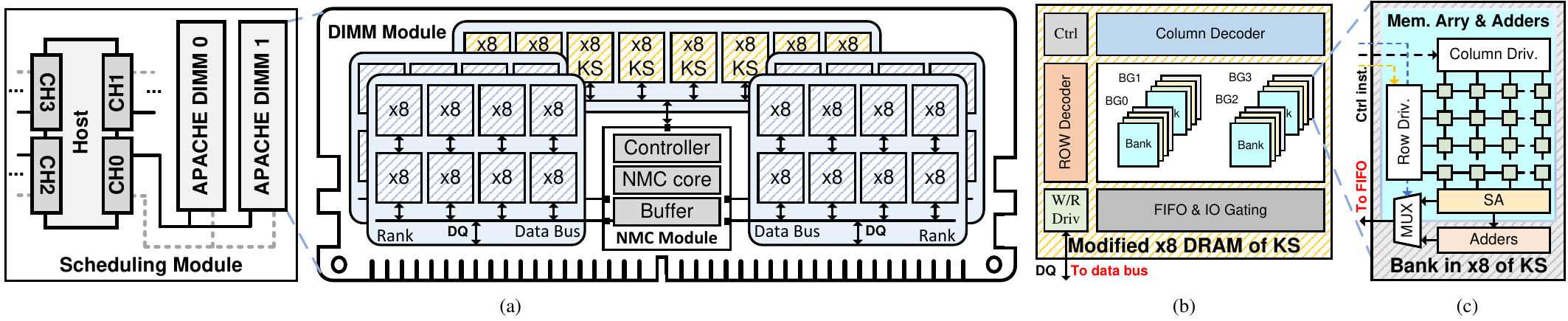}
    \vspace{-18 pt}
    \caption{(a) Overview of \NAME structure; (b) Modified $\times$8 DRAM chip with a hierarchy of array-to-bank-to-bank group; and (c) Modified memory array for computing $\mathsf{KS}$. Dashed and solid lines represent the control and data flows, respectively.}
    \label{fig:overview}
    \vspace{-10 pt}
\end{figure*}

\section{Background}
\label{section: Background and Motivation}

\subsection{Decomposition and Classification of FHE Operators}
\label{subsection: Homomorphic Operators}
Similar to existing work~\cite{HE3DA}, we divide existing FHE schemes into
two primary categories: BFV/CKKS-like and TFHE-like. 
BFV/CKKS-like arithmetic FHE operations include homomorphic addition ($\mathsf{HAdd}$), key switching ($\keyswith$), homomorphic rotation ($\mathsf{HRot}$), homomorphic multiplication ($\mathsf{HMult}$), and CKKS bootstrapping. 
$\mathsf{HMult}$ supports only a finite number of multiplications. Thus, $\mathsf{bootstrapping}$ is designed to restore the multiplicative level budgets.
In both BFV and CKKS, $\mathsf{bootstrapping}$ is the most intricate and resource intensive operation, involving numerous $\keyswith$, $\mathsf{HRot}$, $\mathsf{HMult}$, and $\mathsf{HAdd}$ operations. 
In TFHE, the most basic homomorphic operators are the controlled multiplexer ($\mathsf{CMUX}$) and two key switching types: public ($\mathsf{PubKS}$) and private ($\mathsf{PrivKS}$). Combining TFHE $\mathsf{bootstrapping}$ (mainly hundreds of $\mathsf{CMUX}$) and $\mathsf{PubKS}$ allows various logic gates ($\mathsf{HomGate}$), like $\mathsf{HomAND}$ and $\mathsf{HomOR}$. By jointly using THE $\mathsf{bootstrapping}$ and $\mathsf{PrivKS}$, a circuit bootstrapping (CB) functionality can be achieved to provide higher precision and enhanced flexibility to build advanced applications \cite{HE3DA,VSP}. For more details, see \cite{ARK,F1,HE3DA}.

As shown in \cref{tab:Classification_OP}, we break down all the above FHE operators into lower-level operators, such as the (inverse) number-theoretic transform ($\mathsf{(I)NTT}$), modular addition ($\mathsf{MAdd}$), modular multiplication ($\mathsf{MMult}$), and coefficient automorphism. Based on computation demands and data volume, we further categorize these homomorphic operators into data-intensive and computation-intensive operators. From this table, three significant observations emerge. First, data-intensive operators (like $\mathsf{PrivKS}$ and $\mathsf{PubKS}$) exhibit a shallow computational circuit depth. Consequently, such operations that require less computation but more memory can be positioned nearer to (or integrated within) the memory die to minimize I/O bandwidth usage. Second, several FHE operators (such as CB, $\mathsf{HomGate}$, and CKKS bootstrapping) utilize the same underlying operators but differ in operand bitwidths and the sizes of evaluation keys. Hence, configurable functional units are necessary to support these parameter configurations and FHE operands. Lastly, executing certain FHE operations may cause idle for some lower-level operators, leading to under-utilization of hardware resources of FHE accelerators. Therefore, a versatile interconnect topology for functional units is essential to effectively allocate hardware resources.

\subsection{Circuit Designs for FHE Acceleration}
Recently, FPGA-based FHE accelerators \cite{Poseidon,HEAP,FPT,Cheetah_HPCA2021,HEAX,HPCA2000} have been proposed and provided a speedup of hundreds of times over CPU. Early ASIC accelerators are tailored for either BFV/CKKS-like FHE \cite{ARK,BTS,CraterLake,SHARP_ISCA2023} or TFHE-like FHE \cite{MATCHA,Strix,Morphling}, achieving thousands of performance improvements compared to CPU. More recently, \cite{Alchemist} designed configurable 36-bit functional units to accommodate BFV/CKKS-like FHE operators and $\mathsf{HomGate}$ of TFHE. In addition, processing-in/near-memory accelerators \cite{INSPIRE,Poseidon-NDP,Invited_DAC2021,MemFHE,TVLSI2020,TC2023,IISWC} have also been devised to address the memory issue of FHE evaluation. However, all the above approaches have not discussed dividing and conquering FHE operators as data-heavy and computation-heavy traits, thus limiting compatibility with multi-scheme FHE or not fully aligning with the three core design principles, as \cref{table:compaBetacc} shows. 



\section{Architecture of \NAME}
\label{section: Architecture}

As \cref{fig:overview}(a) shows, \NAME consists of two main modules: the scheduling module and the DIMM module. Upon receiving an FHE program, the host CPU extracts control and data flow graphs of FHE operators. The scheduler then breaks down the data dependency between operators and creates a task queue, where task-level operator parallelism is extracted to exploit computational capabilities of multi-channel DRAM modules. Next, operators are fed into the DRAM module, where most of the data buffering and task execution happens. Lastly, the results of each operator are stored naturally within the DRAM module for subsequent computations. Hence, the overall goal of \NAME is to significantly reduce the off-chip I/O loads for multi-scheme FHE acceleration by properly allocating functional units into the DRAM module. 

\subsection{Architectural Overview}

To carry out complex FHE operations inside the DIMM, computing and storage elements are carefully organized to form three memory levels: the external I/O level, the near-memory computing level, and the in-memory computing level. 

{\bf{\ding{192} I/O Level}}: At the I/O level, we expose the set of operators that constitute the application programming interface (API) of \NAME, including all FHE operators described in \cref{subsection: Homomorphic Operators}. When executing an FHE operator, we assume that all ciphertext data and evaluation keys are pre-loaded into the DIMMs, which we believe is a reasonable assumption since it also applies to the regular computation model.

{\bf{\ding{193} Near-Memory Computing Level}}: The NMC level houses most low-level FUs with main components: i) standard memory chips, ii) the NMC module, and iii) the modified in-memory computing module. We use standard $\times$8 DRAM chips with unchanged internal architectures but individually connecting each rank data bus to data buffers in the NMC module. By parallelizing multiple DRAM ranks, \NAME provides substantial internal bandwidth to the NMC module, so that excessive I/O communication can be avoided. 
Inside the NMC module, we incorporate configurable FUs implementing the respective underlying operators, and the FUs are structured with a flexible interconnect topology to efficiently accommodate various data flows of FHE operators.

{\bf{\ding{194} In-Memory Computing Level}}: As sketched in \cref{fig:overview}(b) and \ref{fig:overview}(c), we place application-specific circuits on the in-memory computing level to reduce the bandwidth bottleneck introduced by the TFHE operators of $\privks$ and $\pubks$.
As \cref{tab:Classification_OP} shows, the required on-chip key sizes is extremely large for $\privks$ and $\pubks$, while the computation circuit is shallow (only a couple of adders). Specifically, both operators are calculated by \cref{eq:PuBKS} and \cref{eq:PrivKS} \cite{tfhe}, 
\begin{align}
    &\mathsf{PubKS} (f,\mathbf{KS},\mathbf{c}) = (0,f(b^{1},\cdots,b^{p})) - \sum_{i=1}^{n}\sum_{j=1}^{t} \hat{a}_{i,j} \cdot \mathbf{KS}_{i,j}, \label{eq:PuBKS}\\
    &\mathsf{PrivKS} (\mathbf{KS}^{(f)},\mathbf{c}) = - \sum_{z=1}^{p}\sum_{i = 1}^{n+1}\sum_{j=1}^{t}  \hat{c}^{(z)}_{i,j} \cdot \mathbf{KS}^{(f)}_{z,i,j}, \label{eq:PrivKS}
\end{align}
where $\mathbf{c}$ is the set of $p$ FHE ciphertexts, and $f$ is a Lipschitz continuous function. $\hat{a}_{i,j}$ is the $j^{th}$ bit of $a_i$, while $\hat{c}^{(z)}_{i,j}$ is the $j^{th}$ bit of the $i^{th}$ coefficient in the $z^{th}$ ciphertext. $\mathbf{KS}$ and $\mathbf{KS}^{(f)}$ are the key-switching keys. 
We insert accumulation adders at the bank level of the $\times$ 8 DRAM chip and preload the evaluation keys. The NMC module only needs to transmit $n \times t$ and $p \times (n+1) \times t$ bits to modified $\times$8 DRAM chips for $\hat{a}_{i,j}$ of $\mathsf{PubKS}$ and $\hat{c}^{(z)}_{i,j}$ of $\mathsf{PrivKS}$, respectively. In contrast, existing FHE accelerators require loading the heavy $\mathsf{KS}$ key from memory to computing modules.

\begin{figure}[t!]
    \centering
    \includegraphics[width = 76mm]{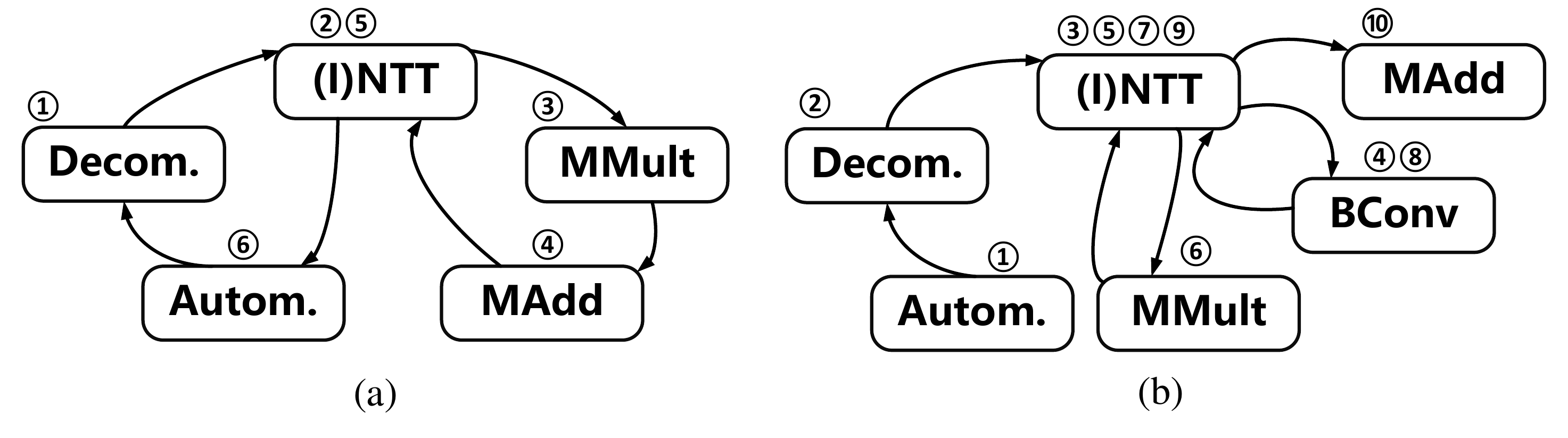}
    \vspace{-12 pt}
    \caption{Dataflow of (a) $\mathsf{CMUX}$ and (b) $\hrot$ and $\cmult$ (2 to 9). $\bconv$ consists of $\mmult$ and $\madd$.} 
    \label{fig:dataflow_CKKS_TFHE}
    \vspace{-8 pt}
\end{figure}

\begin{figure}[t!]
    \centering
    \includegraphics[width = 76mm]{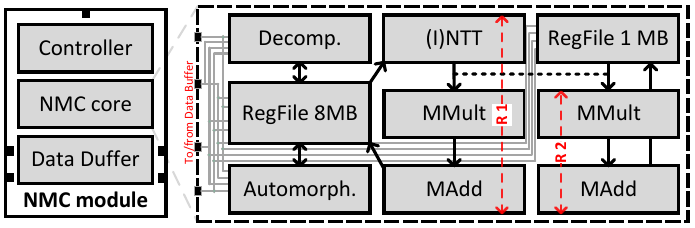}
    \vspace{-8 pt}
    \caption{The topology of NMC module. Dashed line stands for wires with transistors to control whether to link $\mathsf{(I)NTT}$ FU with $\mathsf{MMult}$ FU.}
    \label{fig:The topology of NMC module}
    \vspace{-8 pt}
\end{figure}

\subsection{Task-level Scheduling and Parallelism}
\label{section: Scheduling Data}


The design of an efficient and effective scheduling strategy is essential to exploit the parallelism capability of multi-channel DIMMs. Thus, the mapping and aggregation intermediate results have to be carefully treated to avoid excessive inter-DIMM communication. For such a purpose, we define scheme-specific scheduling strategies at the FHE application level, i.e., task-level. First, we consider two basic cases of FHE tasks: those without a specific ciphertext sequence and those with. More complex tasks can be constructed by combining the two basic cases. Without explicit data dependency, ciphertexts are processed concurrently on multiple \NAME DIMMs. Only the local result is then transmitted to a DIMM via the host CPU for aggregation. In tasks with specific order, if operands fit within a single DIMM's capacity, the task is processed locally. Otherwise, the local result is passed to the next DIMM for subsequent evaluations. 

To ensure that ciphertext results are efficiently transferred between DIMMs by host, we need to dynamically select appropriate ciphertext formats on the FHE application layer to minimize transmission latency. The multi-scheme FHE operates mainly on two ciphertext formats: the $\lwe$ and $\rlwe$ ciphertexts. A batch of $\lwe$ ciphertexts can be packed into $\rlwe$ ciphertexts for storage and transfer \cite{HE3DA}. 
In \NAME, when \cref{eq:pack} holds, we pack $t$ $\lwe$ ciphertexts into an $\rlwe$ for efficient transfer across DIMMs,  
\begin{align}
    \mathbf{T}_{\rm{Pack(t\cdot \mathsf{LWE})}} + \rm{\rlwe}. \mathbf{T}_{\rm{Transfer}} \leq t \cdot \rm{\lwe}.\mathbf{T}_{\rm{Transfer}},
    \label{eq:pack}
\end{align}
where $\mathbf{T}_{\rm{Pack}}$ and $\mathbf{T}_{\rm{Transfer}}$ represent the latencies of packing and transfer, respectively.

\section{Design Exploration of Functional Units}
\label{section: Functional Unit}

To design a near-computing module for multi-scheme FHE, we encounter two major design challenges. First, as illustrated in \cref{fig:dataflow_CKKS_TFHE}(a) and (b), TFHE-like and BFV/CKKS-like schemes exhibit distinct data and control flows. 
Thus, regardless of which scheme we use as the reference for the NMC design, there will always be inefficiencies when the other scheme is used. 
Second, within BFV/CKKS-like schemes, many tasks adopt only $\hadd$ and $\pmult$ without invoking NTT. As a result, if the NTT, $\mmult$, and $\madd$ circuits are pipeline together, the NTT modules become idle when only $\hadd$ and $\pmult$ are called. To tackle the above challenges, we propose a flexible NMC module with configurabilities on both the interconnect and FU levels.

\subsection{The NMC Module and its Interconnection}

As outlined in \cref{fig:The topology of NMC module}, the NMC module consists of three parts: the interconnect controller, the NMC core, and the data buffer. During the execution of an FHE operator, data is loaded into the buffer. Then, the controller configures the data path in the NMC core for computations. The NMC core includes six FUs: register file, decomposition unit $\decomp$, (I)NTT unit $\nintt$, modular multiplication $\mmult$, modular addition $\madd$, and automorphism unit $\auto$. The $\nintt$ supports NTT and INTT by using different twiddle factors.

While the designs of the FUs themselves are close to those of the existing works~\cite{F1,CraterLake,ARK,Strix}, the interconnects need to be completely redesigned due to the complex arithmetic characteristics of multi-scheme FHE operators. We observe that the $\nintt$-$\mmult$-$\madd$ routine is a frequently used control flow for BFV/CKKS and TFHE, while automorphism and decomposition operators are comparatively less used and require less bandwidth, as shown in \cref{fig:dataflow_CKKS_TFHE} (a) and (b). Based on this observation, we propose a configurable interconnect topology for NMC modules, featuring two pipelines: $\nintt$-$\mmult$-$\madd$ and $\mmult$-$\madd$, as depicted in \cref{fig:The topology of NMC module}. The former routine, coupled with a central 8MB register file, can execute operators, like $\mathsf{CMUX}$ of $\mathsf{HomGate}$ and $\mathsf{CMult}$. The latter handles $\mathsf{HAdd}$ and $\mathsf{PMult}$ without interrupting the $\nintt$ pipeline. 
We formulate the $\nintt$ FU utilization rate for a single fixed pipeline and our configurable topology via \cref{equ:one pipeline} and \cref{equ:ours}, respectively.
\begin{align}
    &\mathbf{Utl}_{\rm{NTT}} = \frac{\mathbf{T}_{\rm{ALL}}-\mathbf{T}_{\rm{nonNTT}}}{\mathbf{T}_{\rm{ALL}}}, \label{equ:one pipeline} \\
    &\mathbf{Utl}'_{\rm{NTT}} = \frac{\rm{R1}.\mathbf{T}_{\rm{ALL}} - \rm{R1}.\mathbf{T}_{\rm{nonNTT}}}{\rm{R1}.\mathbf{T}_{\rm{ALL}} \bigcup \rm{R2}.\mathbf{T}_{\rm{ALL}}}, \label{equ:ours}
\end{align}
where $\mathbf{T}_{\rm{ALL}}$ stands for the overall latency of a routine, and $\rm{R1}.\mathbf{T}_{\rm{ALL}} \bigcup \rm{R2}.\mathbf{T}_{\rm{ALL}}$ is the union of timing segments of routines $\rm{R1}$ and $\rm{R2}$. The union run-time is not larger than $\mathbf{T}_{\rm{ALL}}$. Meanwhile, it holds $\rm{R1}.\mathbf{T}_{\rm{nonNTT}} \geq 0.5 \times \mathbf{T}_{\rm{nonNTT}}$ (resp. $\rm{R1}.\mathbf{T}_{\rm{nonNTT}} = $ $\mathbf{T}_{\rm{nonNTT}}$) for $\hadd$/$\pmult$ (resp. $\mathsf{CMUX}$). Thus, we can observe a significant improvement in the utilization rate of the $\nintt$ FU for both CKKS/BFV-like and TFHE-like schemes.

\begin{figure}[t]
    \centering
    \includegraphics[width = 88mm]{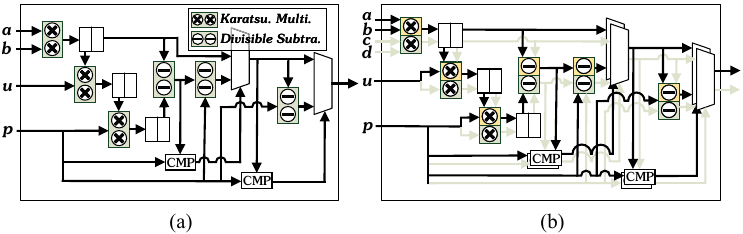}
    \vspace{-22 pt}
    \caption{The proposed configurable modular multiplier, (a) working as a 64-bit modular multiplier, and (b) working as two parallel 32-bit modular multiplier.}
    \label{fig:configurable MM}
    \vspace{-8 pt}
\end{figure}

\begin{figure} [t]
    \centering
    \includegraphics[width=83 mm]{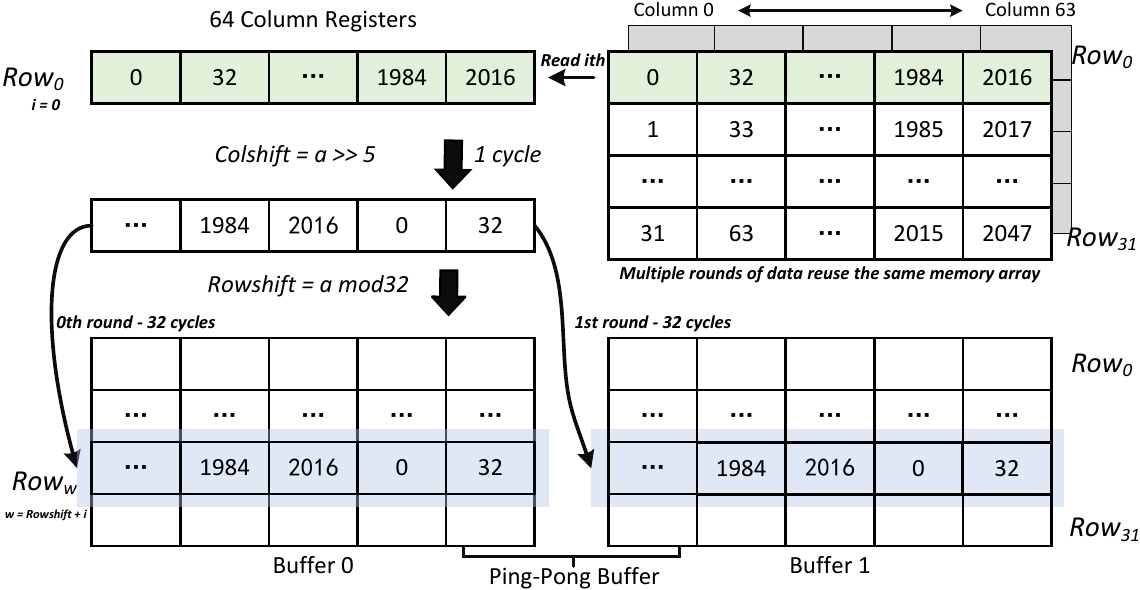}
    \vspace{-9 pt}
    \caption{The structure of automorphism module.}
    \label{fig:Automorphism}
    \vspace{-13 pt}
\end{figure}

\subsection{FU Configuration in the NMC Module}
As illustrated in Table II, a general-purpose FHE accelerator must accommodate 
operands of 32 and 64 bits. 
To accommodate the pivot bit-widths, we have designed the FUs with configurable bitwidth for our \NAME. 

{\bf{(1) Configurable $\mathsf{MMult}$ and $\madd$ circuits (\cref{fig:configurable MM})}}: The proposed configurable structure of $\mmult$ is shown in \cref{fig:configurable MM}. First, a 64-bit adder can be split into two 32-bit adders by cutting at the carry bit. Meanwhile, our key observation is that the Karatsuba multiplier handles a $k$-bit multiplication with three $\frac{k}{2}$-bit multipliers. By modifying the structure of a 64-bit Karatsuba multiplier, it can perform 64-bit or at least two parallel 32-bit multiplications. Then, we can use three $k$-bit multipliers and subtractors to build a $k$-bit $\mmult$. This allows arranging our 64-bit adders and multiplier to create the 64-bit configurable $\mmult$ as in \cref{fig:configurable MM}. Lastly, we can simply use the 64-bit configurable adder and subtractor to implement a configurable $\madd$ FU.


\textbf{(2) Configurable NTT FU}: In this work, we use our configurable $\mmult$ and $\madd$ as basic blocks to build the fully-pipelined four-step NTT FU, which is widely adopted for FHE acceleration \cite{F1,ARK,CraterLake}. By such an approach, the (I)NTT circuit can be flexibly configured into either one high bit-width (up to 64 bits) (I)NTT FU or two parallel low bit-width (each supporting up to 32 bits) (I)NTT FUs.

\textbf{(3) Automorphism (\cref{fig:Automorphism})}: In addition to supporting multiple operand bitwidths, the automorphism unit must handle distinct coefficient rotation rules in CKKS and TFHE. \cite{F1} shows BGV and CKKS automorphism is effectively implemented by four permute/transpose SRAM buffers.
It is observed that the TFHE automorphism can be implemented using only a few shift registers and a single SRAM buffer, highlighting an inevitable disparity between the automorphism operators employed in CKKS and TFHE.

Our key insight is that, in the TFHE bootstrapping, rotating a ciphertext is implemented by calculating $\mathbf{ACC} \leftarrow X^{-\mathbf{a}_i} \cdot \mathbf{ACC} - \mathbf{ACC}$, where $\mathbf{ACC}$ is initialized as a ciphertext, and $X^{-\mathbf{a}_i}$ indicates the rotation index. To facilitate the pipeline-style update procedure for ACC, we require an SRAM buffer to store the original polynomial, as well as a ping-pong SRAM buffer for the coefficient rotation of $X^{-\mathbf{a}_i} \cdot \mathbf{ACC}$. Thus, we propose the integration of the coefficient rotation and $- \mathbf{ACC}$ together, which ensures three SRAM buffers available for the automorphism of CKKS and TFHE, as \cref{fig:Automorphism}. Furthermore, in the working state of CKKS, data are read from the third SRAM buffer following a predetermined order to achieve the equivalent effect of the last-time transpose operation.

\begin{table}[t]
    \footnotesize
    \centering
    \caption{\NAME DIMM configuration.}
    \vspace{-8 pt}
    \begin{tabular}{cc|cc}
    \toprule
         Memory capacity& 8 GB &DRAM chip& 3200 MT/s\\
         Rank per DIMM&8&Clock frequency&1600 MHz\\
         NMC per DIMM&1&tRCD-tCAS-tRP&22-22-22\\
    \bottomrule
    \end{tabular}
    \label{tab:my_label}
    \vspace{-8 pt}
\end{table}

\begin{table}[t]
    \footnotesize
    \centering
    \caption{Area and Thermal Design Power (TDP) of NMC moduler, with breakdown by component.}
    \vspace{-8 pt}
    \begin{tabular}{lrr}
    \toprule
      Component&Area [mm$^2$]&Power [W]\\
      \midrule
      64-point (I)NTT $\times$ 4 & 13.04 & 6.28 \\
      Automorphism $\times$ 2 & 2.4 & 0.6 \\
      Decomposition $\times$ 2 & 0.03 & 0.02 \\
      Modular Multiplier $\times$ 256 $\times$ 2 & 5.0 & 3.01 \\
      Modular Adder $\times$ 256 $\times$ 2 & 0.36 & 0.39 \\
      Adders in each $\times$8 DRAM & 0.12 & 0.02\\
      \midrule
      Regfile (8 + 1 MB) & 14.4 & 1.01 \\
      Data Buffer (24 MB) & 38.4 & 2.7 \\
      \midrule
      Total NMC module & 73.75 & 14.03\\
      \bottomrule
    \end{tabular}
    \label{tab:hardware overhead}
    \vspace{-11 pt}
\end{table}

\section{Implementation and Evaluation}
\label{section: Implementation}

\subsection{Implementation and Setup}


To evaluate the performance of our \NAME, we have developed a cycle-accurate behavioral simulator based on the TFHE and CKKS schemes. For simulating the buffer and DRAM behaviors in each DIMM, we have used NVsim~\cite{NVSim} and CACTI~\cite{CACTI}, which are widely-used tools for memory analysis. The latency, area, and power characteristics of the NMC module are estimated using the Synopsys Design Compiler at a frequency of 1 GHz. The cache, memory, and function units are evaluated at a 22 nm technology node. The DIMM configuration and hardware overhead are presented as \cref{tab:my_label} and \cref{tab:hardware overhead}, respectively.


\begin{table}[t]
    \footnotesize
    \centering
    \caption{Comparison of CKKS operators is with $N = 10^{16}$ and $L =$ 44. HomGate and Circuit Bootstrapping are set to 80-bit and 128-bit security, respectively. We adopt ``operators per second'' (op./s) as the metric.}
    \vspace{-8 pt}
    \begin{tabular}{l|lll|ll}
    \toprule
      Accelerators & $\mathsf{HAdd}$ & $\mathsf{CMult}$ & $\mathsf{HRot}$ & $\mathsf{HomGate}$ & $\mathsf{CBoot.}$\\
    \midrule    
     Poseidon \cite{Poseidon} & 13.3K & 273 & 302 & - & -\\
     F1 \cite{F1} & N/A & N/A & N/A & - & -\\
     Strix \cite{Strix} &-&-&- & 74.7K & $\leq$ 2.6K\\
     Morhpling \cite{Morphling} &-&-&-&147K & $\leq$ 7.4K\\
     Alchemist \cite{Alchemist}& 710K & 7.14K & 7.18K & 123K & -\\
     \NAME $\times$ 2 & 355K & 7.4K & 7.72K & 500K & 49.6K\\
     \NAME $\times$ 4 & 708K& 17.7K & 15.4K &1,000K & 99.2K\\
     \bottomrule
    \end{tabular}
    \begin{tablenotes}
     \item[] 
     We use N/A to denote an accelerator that supports a specific FHE operator but without reported performance, and utilize - to mark the absence of support for an FHE operator.
    \end{tablenotes}
    \label{tab:comparison}
    \vspace{-8 pt}
\end{table}

\begin{figure*}[h]
    \centering
    \includegraphics[width = \linewidth]{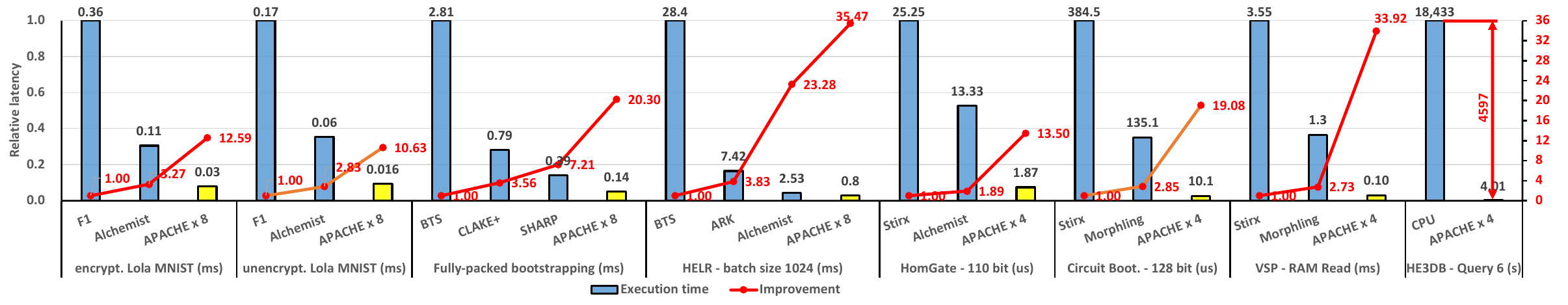}
    \vspace{-23 pt}
    \caption{Performance comparison of \NAME with current FHE accelerators. To ensure similar area overhead with the comparison accelerators, we report the performance of \NAME $\times$ 4 and $\times$ 8 for TFHE and CKKS benchmarks, respectively.}
    \label{fig:Application_Comp}
    \vspace{-11 pt}
\end{figure*}

\subsection{Performance and Comparison} 
\label{subsec:Performance and Comparison}

\begin{figure}[t]
    \centering
    \includegraphics[width = 83 mm]{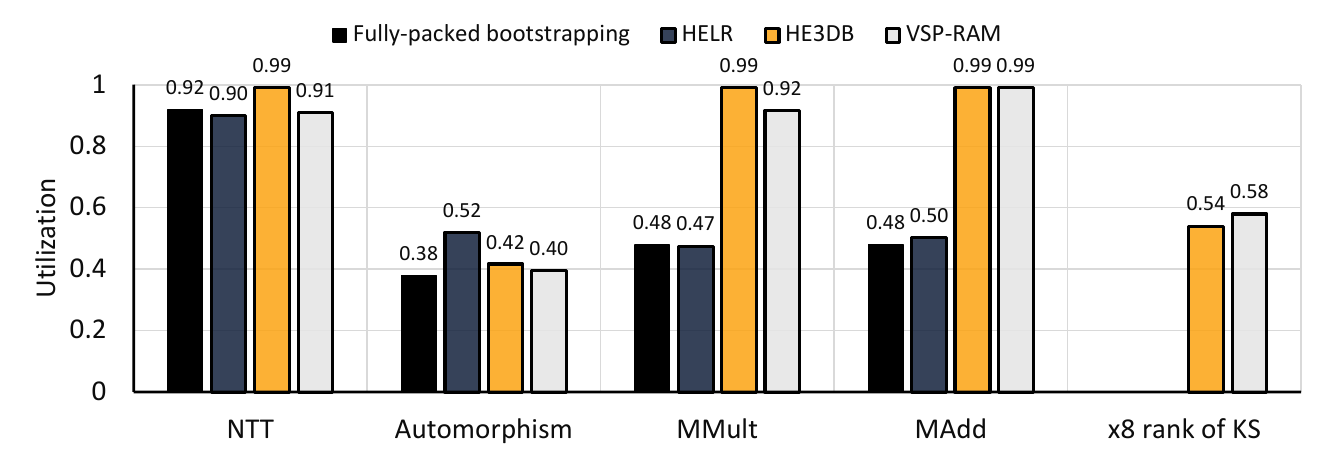}
    \vspace{-13 pt}
    \caption{Resource utilization of \NAME.}
    \label{fig:utilization}
    \vspace{-14 pt}
\end{figure}

For comparison purposes, we evaluate \NAME with the state-of-the-art FHE accelerators \cite{Poseidon,F1,CraterLake,ARK,BTS,Strix,Morphling,SHARP_ISCA2023,Alchemist}. Meanwhile, we evaluate the performance of the CB operator for Strix \cite{Strix} and Morphling \cite{Morphling}  with each max bandwidth available for $\mathsf{PrivKS}$. To facilitate evaluation and comparison, we incorporate various CKKS and TFHE operators, along with five distinct FHE applications, such as Lola MNIST \cite{Lola}, fully-packed bootstrapping \cite{TFHEBoot2018}, HELR \cite{LR_AAAI2019}, VSP \cite{VSP}, and HE$^3$BD \cite{HE3DA}. 
All assessments adhere to \cite{Alchemist}.

\textbf{Performance of FHE Operators}: 
\cref{tab:comparison} shows that our \NAME architecture achieves significant advantages over comparable solutions when executing TFHE operators. Even with only two parallel \NAME DIMMs, we achieve remarkable performance, surpassing the recent work of Strix, Morphling, and Alchemist by 6.69$\times$, 6.75$\times$, and 4.07$\times$, respectively. 
When implementing circuit bootstrapping, our \NAME attains the 19.08$\times$ and 6.7$\times$ speedup than Strix and Morphling, respectively. This significant edge is primarily because we decouple $\mathsf{PrivKS}$ and $\mathsf{PubKS}$ from computation clusters and process both within specific $\times$8 ranks, reducing the bandwidth load by 3.15 $\times$ 10$^{5}$ times and 3.05 $\times$ 10$^{4}$ times, respectively. In contrast, Strix (resp. Morphling) takes around 24\,ms (resp. 7.7\,ms) for loading the $\mathsf{PrivKS}$ key of 1.8\,GB for circuit bootstrapping. In addition, \cite{Alchemist} relies on 36-bit circuits, which is incompatible with 64-bit circuit bootstrapping.

\textbf{Full-system Performance}: 
As \cref{fig:Application_Comp} shows, \NAME with 8 DIMMs achieves a speedup of 12.59$\times$ and 10.63$\times$ for Lola MNIST \cite{Lola} with encrypted and unencrypted weights, respectively. Moreover, our \NAME DIMM $\times$ 8 is 20.3$\times$ and 35.47$\times$ faster than the baseline of BTS for the CKKS schemes of fully-packed bootstrapping \cite{TFHEBoot2018} and HELR \cite{LR_AAAI2019}, respectively. 
Such improvement APACHE has achieved is primarily attributed to its flexible interconnect topology that supports fine-grained operator scheduling. Concretely, such real applications of CKKS commonly include a large number of $\mathsf{PMult}$ and $\mathsf{HAdd}$ independent of NTT. Our \NAME allows both operators to be (partially or totally) executed in pipeline routine 2. Thus, our APACHE significantly improves the utilization of NTT FUs and overall performance. When \NAME performs the TFHE-based VSP \cite{VSP}, we observe 33.92$\times$ and 12.4$\times$ speedup against Strix and Morphling, respectively. The VSP relies on expensive circuit bootstrapping to generate ciphertext addresses. Due to the optimization mismatch in CB over Strix and Morhpling, \NAME exhibits more competitive performance in VSP. Furthermore, \cref{fig:Application_Comp} shows that our \NAME supports HE$^3$DB \cite{HE3DA} based on both TFHE and CKKS schemes, suppressing CPU by 4597$\times$.

\subsection{Architectural Analysis}



\textbf{Utilization Rates of FUs}: \cref{fig:utilization} demonstrates that our \NAME has a high utilization of hardware resources. The utilization of the $\nintt$ FU always stays above 90\%, while that of the existing accelerators only ranges from 50\% to 85\%. It indicates that our interconnect topology can effectively reduce the operating time of NTT-independent operations. 
In addition, the utilization of the in-memory KS module is around 50\%, which means that the multi-level storage architecture effectively reduces the bandwidth caused by loading KS keys.

\textbf{Remark on Data Communication across DIMMs}: Our APAHCE finds the aggregation point to minimize the data volume that needs to be transferred. Thus, the time required for data propagation across DIMMs is substantially smaller than that needed by local computations. For example, a APACHE DIMM needs 0.38 ms to read a $\lwe$ ciphertext from the VSP RAM, in which 512 $\lwe$ ciphertexts are stored. By contrast, forwarding the readout $\lwe$ cipher to another DIMM via the host only requires 0.31 us with given 32GB/s of I/O bandwidth. That is, our APACHE can cover the communication latency within the computation duration.

\section{Conclusion}
\label{section: Conclusion}
In this paper, we present a near-memory computing architecture to accelerate both BFV/CKKS-like and TFHE-like FHE schemes. By classifying operators of both FHE lanes into data-heavy and computation-heavy types, we emphasize the importance of allocating FUs at the correct memory level, over simply increasing computing resources, for effectively accelerating multi-scheme FHE. 
We propose \NAME that allocates fully pipelined circuits at both near-memory and in-memory levels with configurable FU topology to enhance hardware utilization. Experiments show that \NAME can reduce the external I/O bandwidth by as much as 3.15 $\times$ $10^{5}$. Moreover, \NAME behaves better than current ASIC accelerators in multiple FHE applications.

\newpage
\bibliographystyle{IEEEtranS}
\bibliography{IEEEabrv,refs}

\end{document}